\documentclass[aps,prc,twocolumn,10pt,superscriptaddress]{revtex4-1}

\usepackage{hyperref}
\usepackage{graphicx}
\usepackage{amsmath}
\usepackage{slashed} 
\usepackage[dvipsnames,svgnames,table]{xcolor}

\hypersetup{
  colorlinks,
  unicode=True,
  linkcolor=Blue,
  citecolor=Blue,
  urlcolor=Blue,
  pdftitle={Exploring jet transport coefficients in the strongly interacting quark-gluon plasma},
  pdfauthor={Ilia Grishmanovskii},
}

\graphicspath{ {figures/} }

\newcommand{\qhat}{\hat{q}}
\newcommand{\dpi}{(2\pi)}
\newcommand{\calO}{\mathcal{O}}
\newcommand{\Msq}{|\bar{\mathcal{M}}|^2}
\renewcommand{\vec}{\mathbf}

\begin{document}


\title{Exploring jet transport coefficients by elastic scattering in the strongly interacting quark-gluon plasma}


\author{Ilia Grishmanovskii}
\email{grishm@itp.uni-frankfurt.de}
\affiliation{Institut f\"ur Theoretische Physik, Johann Wolfgang Goethe-Universit\"at,Max-von-Laue-Str.\ 1, D-60438 Frankfurt am Main, Germany}

\author{Taesoo Song}
\affiliation{GSI Helmholtzzentrum f\"ur Schwerionenforschung GmbH,Planckstrasse 1, D-64291 Darmstadt, Germany}

\author{Olga Soloveva}
\affiliation{Helmholtz Research Academy Hesse for FAIR (HFHF), GSI Helmholtz Center for Heavy Ion Physics, Campus Frankfurt, 60438 Frankfurt, Germany}
\affiliation{Institut f\"ur Theoretische Physik, Johann Wolfgang Goethe-Universit\"at,Max-von-Laue-Str.\ 1, D-60438 Frankfurt am Main, Germany}

\author{Carsten Greiner}
\affiliation{Institut f\"ur Theoretische Physik, Johann Wolfgang Goethe-Universit\"at,Max-von-Laue-Str.\ 1, D-60438 Frankfurt am Main, Germany}
\affiliation{Helmholtz Research Academy Hesse for FAIR (HFHF), GSI Helmholtz Center for Heavy Ion Physics, Campus Frankfurt, 60438 Frankfurt, Germany}

\author{Elena Bratkovskaya}
\affiliation{GSI Helmholtzzentrum f\"ur Schwerionenforschung GmbH,Planckstrasse 1, D-64291 Darmstadt, Germany}
\affiliation{Institut f\"ur Theoretische Physik, Johann Wolfgang Goethe-Universit\"at,Max-von-Laue-Str.\ 1, D-60438 Frankfurt am Main, Germany}
\affiliation{Helmholtz Research Academy Hesse for FAIR (HFHF), GSI Helmholtz Center for Heavy Ion Physics, Campus Frankfurt, 60438 Frankfurt, Germany}


\date{\today}
\begin{abstract}
We study the interaction of leading jet partons in a strongly interacting quark-gluon plasma (sQGP) medium based on the effective dynamical quasi-particle model (DQPM). The DQPM describes the non-perturbative nature of the sQGP at finite temperature $T$ and baryon chemical potential $\mu_B$ based on a propagator representation of massive off-shell partons (quarks and gluons) whose properties (characterized by spectral functions with $T,\mu_B$ dependent masses and widths) are adjusted to reproduce the lQCD EoS for the QGP in thermodynamic equilibrium. We present the results for the jet transport coefficients, i.e. the transverse momentum transfer squared per unit length $\qhat$ as well as the energy loss per unit length $\Delta E =dE/dx$, in the QGP and investigate their dependence on the temperature $T$ and baryon chemical potential $\mu_B$ as well as on jet properties such as the leading jet parton momentum, mass, flavor, and the choice of the strong coupling constant. In this first study only elastic scattering processes of a leading jet parton with the sQGP partons are explored discarding presently the radiative processes (such as gluon Bremsstrahlung). We present a comparison of our results for the elastic energy loss in the sQGP medium with the pQCD results obtained by the BAMPS and LBT models as well as with other theoretical approaches such as lattice QCD and the LO-HTL and also with estimates of $\qhat/T^3$ by the color string percolation model (CSPM) and the JET and JETSCAPE Collaborations based on a comparison of hydrodynamical calculations with experimental heavy-ion data.
\end{abstract}


\maketitle


\section{Introduction}

Ultra-relativistic heavy-ion collisions performed at the Super Proton Synchrotron (SPS), the Large Hadron Collider (LHC) at CERN and the Relativistic Heavy-Ion Collider (RHIC) provide an access to a new hot and dense phase of matter, the quark-gluon plasma (QGP). Although at high momentum transfer the QCD describes quarks and gluons as weakly-coupled, the experimental observations \cite{BRAHMS:2004adc,PHENIX:2004vcz,PHOBOS:2004zne,STAR:2005gfr} suggest that the matter created in central heavy-ion collisions shows the properties of a strongly-coupled liquid \cite{Shuryak:2004cy,Gyulassy:2004zy}. An understanding of the properties of the QGP is one of the main goals of current research in heavy-ion physics.

One way to study the degrees of freedom of the QGP matter is to use 'external probes' to investigate their attenuation and deflection in the QGP medium. However, the very short lifetime of the QGP does not allow to study it with a literally external probe, but high-energy objects created in initial parton-parton scatterings can be used instead. Produced in the early stage of the heavy-ion collisions, these partons get high transverse momentum and traverse the QGP interacting with the medium through collisional and radiative processes. In the end they fragment into a collimated entity of hadrons which is called a jet. The modification of these jets caused by the medium is denoted as 'jet quenching', which has been first observed at RHIC \cite{STAR:2003pjh,STAR:2002svs} and later confirmed at LHC \cite{ALICE:2010yje,ATLAS:2010isq}. It has been measured that the ratio $R_{AA}$ of the hadron high $p_T$ spectra from $A+A$ collisions is substantially suppressed compared to $p+p$ spectra scaled by the number of binary collisions $N_{coll}$ due to the energy loss of jet partons when traversing the medium. This makes jets a sensitive probe to perform a tomographic study of the hot and dense matter produced in heavy-ion collision - from the initial far-off-equilibrium stage to the nearly equilibrated strongly interacting Quark-Gluon Plasma (sQGP), followed by hadronization and hadronic interactions in the final stage. Jets take part in the subsequent in-medium evolution, which leads to a modification of their properties as well as the properties of the medium.

Starting from the pioneering work by Bjorken \cite{Bjorken:1982} in 1982 a substantial progress in the understanding of the jet energy loss has been made in the last decades. Theoretical studies in Refs. \cite{Thoma:1990fm,Dokshitzer:1991fc,Mrowczynski:1991da,Gyulassy:1993hr, Baier:1996kr, Baier:1996sk, Baier:1998kq, Zakharov:1996fv, Gyulassy:1999zd, Wiedemann:2000za, Guo:2000nz, Arnold:2002ja, Armesto:2003jh, Djordjevic:2008iz,Zapp:2008gi,Majumder:2009ge,Caron-Huot:2010qjx} etc. showed that the parton energy loss can be described by a series of jet transport coefficients such as the jet quenching parameter $\qhat$ which denotes the transverse momentum transfer squared per unit length (or time) of the propagating hard parton to the QGP medium as well as the energy loss per unit length $\Delta E =dE/dx$.

Using phenomenological models for the in-medium evolution of HICs (from a simple Bjorken expansion to more sophisticated hydrodynamical calculations), one can relate the theoretically calculated transport coefficients to the experimental observables. Such an extraction of $\qhat$ was performed by the JET \cite{Burke:2013yra} and JETSCAPE \cite{Soltz:20194J} Collaborations by comparing several approaches to parton propagation and energy loss in dense matter with the experimental data for the hadron suppression factor $R_{AA}$. On the other hand the jet quenching can be calculated directly by microscopic transport approaches which incorporate partonic interactions directly in terms of cross sections. Such study of jet attenuation has been performed within the BAMPS (Boltzmann Approach to Multi-Parton Scatterings) \cite{Uphoff:2014cba,Senzel:2020tgf,Fochler:2013epa} based on massless degrees of freedom and interactions of perturbative QCD which incorporated the elastic partonic scattering and radiative process by implementing the improved Gunion-Bertsch matrix element \cite{Gunion:1981qs} for the emission of gluons \cite{Fochler:2013epa} including the Landau-Pomeranchuk-Migdal effects \cite{Landau:1953um,Migdal:1956tc}. The hadronic energy loss has been studied using the microscopic HSD (Hadron-String Dynamics) approach \cite{Cassing:2003sb}. In the both cases the initial jet shower was simulated using the PYTHIA model \cite{Sjostrand:2006za}.
 
The jet energy loss in the QGP medium occurs due to elastic $2\to 2$ partonic scatterings as well as by radiative $2\to 3$ processes with emission of a gluon, i.e. gluon bremsstrahlung. According to the pQCD calculations the radiative energy loss is dominant at high jet momentum despite that the gluon radiation has to be screened in the medium due to the Landau-Pomeranchuk-Migdal (LPM) effect \cite{Landau:1953um,Migdal:1956tc}. Nevertheless, as advocated in Refs. \cite{Mustafa:2003vh,Dutt-Mazumder:2004loa} the elastic energy loss becomes equally important at intermediate and even dominant at low jet momenta. As shown in Ref. \cite{Djordjevic:2006tw} the important role of collisional energy loss survives also for finite size systems as probed in heavy-ion collisions and thus becomes crucial to obtain reliable predictions for jet quenching. The region of low and intermediate jet momentum is very interesting in terms of studying the QCD medium recoil and partial thermalization of low momenta partons from the jet shower. These jet-induced medium excitation have been explored within the Linear Boltzmann transport (LBT) model \cite{He:2015pra} based on pQCD. However, there is a fundamental complexity in employing the pQCD calculations for such studies due to the fact that at low parton momenta ($p < 10$ GeV/c) the non-perturbative phenomena of QCD related to the non-Abelian nature of QCD play a dominant role. Thus, the study of the attenuation of low and intermediate momentum jets requires to go beyond on-shell pQCD. The first estimates of the non-perturbative phenomena of jet energy loss from lattice QCD (lQCD) became available in Refs. \cite{Majumder:2012sh,Panero:2013pla,Kumar:2018cgf,Kumar:20198n,kumar2020jet}. Moreover, this topic has been studied in the framework of finite temperature field theory in the hard thermal loop (HTL) approximation \cite{Djordjevic:2006tw,Djordjevic:2008iz,Djordjevic:2009cr,Djordjevic:2011dd,Caron-Huot:2008zna}, in non-perturbative calculations within EQCD \cite{Moore:2019lgw} and in the AdS/CFT models \cite{Liu:2006ug, Grefa:2022sav} which show the importance of non-perturbative effects for the understanding of the transverse jet momentum broadening and jet quenching, as advocated recently also in Ref. \cite{Moore:2021myb}.

The goal of this study is to evaluate the interaction of jet partons with the strongly interacting non-perturbative QGP medium based on the QCD inspired effective field-theoretical model -- the Dynamical QuasiParticle Model (DQPM) \cite{Peshier:2005pp,Cassing:2007nb,Cassing:2007yg,Berrehrah:2016vzw,Moreau:2019vhw,Soloveva:2019xph}. The DQPM is a 2PI model based on a propagator representation with effective strongly interacting quarks/antiquarks and gluons that have a finite width in their spectral functions (defined by the imaginary parts of the propagators). The properties of quasiparticles -- their masses and widths as well as the effective coupling constant -- depend on the temperature and baryon chemical potential $\mu_B$ and are adjusted by fitting the DQPM entropy density to the lattice QCD (lQCD) data at zero $\mu_B$ \cite{Cassing:2007nb,Cassing:2007yg}. The DQPM allows for a microscopic interpretation of the lattice QCD results for thermodynamic quantities in terms of off-shell effective strongly interacting partonic quasiparticles with broad spectral functions. Moreover, the off-shell partonic interaction cross sections are evaluated based on the leading order scattering diagrams and depend on $T, \mu_B$, the invariant energy of the colliding partons $\sqrt{s}$ as well as the scattering angle \cite{Moreau:2019vhw}. These interaction cross sections have been probed by transport coefficients (correlators) in thermodynamic equilibrium: the shear and bulk viscosity, electric conductivity, magnetic susceptibility as well as the full diffusion matrix \cite{Ozvenchuk:2012kh,Cassing:2013iz,Steinert:2013fza,Moreau:2019vhw,Fotakis:2021diq,Soloveva:2020hpr}. The DQPM has been incorporated in the PHSD approach, based on Kadanoff-Baym off-shell dynamics, for the description of the partonic phase and hadronization (cf. the reviews \cite{Cassing:2008nn,Linnyk:2015rco,Bleicher:2022kcu}), and has led to a successful description of a large variety of experimental data.

We note that the DQPM (as implemented in the PHSD) has been applied for studying the energy loss of heavy quarks in heavy-ion collisions \cite{Song:2015sfa,Song:2015ykw}. There it has been found that the experimental data on the ratio $R_{AA}$ as well as on the elliptic flow $v_2$ for charm mesons with $p_T < 10$ GeV/c can be explained by the collisional energy loss only. On the other hand, the large masses of quasiparticles lead to a more effective energy loss per collision compared to the pQCD results (due to flatter angular cross sections) \cite{Berrehrah:2016vzw,Song:2020tfm}. Moreover, due to the rise of the strong coupling constant in the vicinity of the critical temperature $T_C \simeq 0.156$ GeV \cite{Bazavov:2018mes, Borsanyi:2020fev}, in the DQPM the $\hat q/T^3$ grows with decreasing temperature (i.e. by decreasing the collision energy). We note that similar findings have been obtained within the 1PI quasiparticle models with thermal masses \cite{Scardina:2017ipo,Liu:2021dpm}. For the detailed comparisons of pQCD models to quasiparticle models for the charm transport coefficients and dynamical observables in heavy-ion collisions, we refer the reader to Refs. \cite{Berrehrah:2016vzw,Aichelin:2013mra,Rapp:2018qla,Xu:2018gux,Cao:2018ews}.

In this study we concentrate on the energy loss of jet partons due to elastic scattering in the strongly interacting QGP. The collisional energy loss is expected to be the dominant contribution at low and intermediate jet momenta due to the fact that the radiation of massive gluons has to be suppressed in quasiparticle models contrary to pQCD \footnote{The explicit evaluation of the radiative energy loss within the DQPM model is presently in progress.}. We calculate the jet transport coefficients, i.e. the jet quenching parameter $\qhat$ as well as the energy loss per unit length $\Delta E =dE/dx$, within the DQPM as a function of different variables such as the medium temperature $T$, jet momentum $p$, jet mass $M$, jet flavor, and baryon chemical potential $\mu_B$. Here we consider a symmetric setup of quark chemical potentials $\mu_u=\mu_d=\mu_s=\mu_B/3$. 

For this aim we first compute the elastic off-shell cross sections for the scattering of fast jet partons (quark or gluon) based on the DQPM. Contrary to the pQCD cross sections, which contain an infrared divergence and require a regularization by introduction of the Debye mass, the DQPM cross sections are finite due to the thermal masses and widths in the partonic spectral functions (i.e. propagators). This allows for an explicit propagation of such degrees-of-freedom in a microscopic transport approach as the PHSD (as realized for heavy-quarks) based on Kadanoff-Baym dynamics applicable for strongly interacting systems.

We also consider different options for the coupling constant $\alpha_s$ (i.e. ($T,\mu_B$)-dependent vs. constant), that determines the strength of the interaction between jets and medium partons, and investigate how the $\qhat$ coefficient depends on the coupling $\alpha_s$. Furthermore, we compare our results for the collisional energy loss in the sQGP medium with the pQCD results of the BAMPS and LBT models as well as with available lattice QCD and the LO-HTL results and estimates of $\hat q$ by the JET and JETSCAPE Collaborations and the color string percolation model (CSPM) \cite{Mishra:2021yer}. 

This paper is organized as follows: in Sec. \ref{sec:DQPM} we will provide a brief description of the DQPM and its ingredients. Sec. \ref{sec:tcoef} will be devoted to the calculation of the transport coefficients employing the effective propagators and couplings from the DQPM. In Sec. \ref{sec:results} we will show the evaluation of the $\qhat$ transport coefficient and compare our results with the phenomenological extractions made by JET collaboration and also with recent lQCD and LO-HTL calculations. A summary and outlook will be given in Sec. \ref{sec:summary}.


\section{Dynamical Quasiparticle Model}
\label{sec:DQPM}

The dynamical quasiparticle model (DQPM) \cite{Peshier:2005pp,Cassing:2007nb,Cassing:2007yg,Berrehrah:2016vzw,Moreau:2019vhw,Soloveva:2020hpr} is an effective approach which describes the QGP in terms of strongly interacting quarks and gluons. Their properties are fitted to reproduce lattice QCD calculations in thermal equilibrium and at vanishing chemical potential. In the DQPM the quasi-particles are characterized by dressed propagators with complex self-energies, where the real part of the self-energies is related to dynamically generated thermal masses, while the imaginary part provides information about the lifetime and reaction rates of the partons. The spectral functions are no longer $\delta$-functions, but have the form \cite{Linnyk:2015rco}
\begin{align}
  \rho_{j}(\omega,{\bf p}) &= \frac{\gamma_{j}}{\tilde{E}_j}
  \left(\frac{1}{(\omega-\tilde{E}_j)^2+\gamma^{2}_{j}}
  -\frac{1}{(\omega+\tilde{E}_j)^2+\gamma^{2}_{j}}\right) 
  \nonumber\\
  &\equiv \frac{4\omega\gamma_j}{\left( \omega^2 - \vec{p}^2 - M^2_j \right)^2 + 4\gamma^2_j \omega^2}
  \label{eq:spectral_function}
\end{align}
separately for quarks, antiquarks, and gluons ($j = q,\bar q,g$). Here, $\tilde{E}_{j}^2({\bf p})={\bf p}^2+M_{j}^{2}-\gamma_{j}^{2}$, the widths $\gamma_{j}$ and the masses $M_{j}$ from the DQPM are functions of the temperature $T$ and the chemical potential $\mu$. The spectral function is antisymmetric in $\omega$ and normalized as
\begin{equation}
  \label{eq:spectral_function_norm}
  \int\limits_{-\infty}^{\infty}\frac{d\omega}{2\pi}\
  \omega \ \rho_{j}(\omega,{\bf p})=
  \int\limits_{0}^{\infty} d\omega \frac{\omega}{\pi}\ 
  \rho_{j}(\omega,{\bf p})=1.
\end{equation}

The dynamical quasiparticle mass $M$ is given by the HTL thermal mass in the asymptotic high-momentum regime, i.e., for gluons by \cite{Linnyk:2015rco}
\begin{equation}
  \label{eq:Mg}
  M^2_{g}(T,\mu_q)=\frac{g^2(T,\mu_q)}{6}\left(\left(N_{c}+\frac{1}{2}N_{f}\right)T^2
  +\frac{N_c}{2}\sum_{q}\frac{\mu^{2}_{q}}{\pi^2}\right),
\end{equation}
and for quarks (antiquarks) by
\begin{equation}
  \label{eq:Mq}
  M^2_{q(\bar q)}(T,\mu_q)=\frac{N^{2}_{c}-1}{8N_{c}}g^2(T,\mu_q)\left(T^2+\frac{\mu^{2}_{q}}{\pi^2}\right),
\end{equation}
where $N_{c}\ (=3)$ stands for the number of colors while $N_{f}\ (=3)$ denotes the number of flavors. Eq.~(\ref{eq:Mq}) determines masses for the ($u,d$) quarks, the strange quark has a larger bare mass for controlling the strangeness ratio in the QGP. Empirically, we find $M_s(T,\mu_B)= M_{u/d}(T,\mu_B)+ \Delta M$, where $\Delta M$ = 30 MeV, which has been fixed once in comparison to experimental data. Furthermore, the effective quarks, antiquarks, and gluons in the DQPM have finite widths $\gamma$, which are taken in the form \cite{Linnyk:2015rco}
\begin{equation}
		\gamma_{j}(T,\mu_\mathrm{B}) = \frac{1}{3} C_j \frac{g^2(T,\mu_\mathrm{B})T}{8\pi}\ln\left(\frac{2c_m}{g^2(T,\mu_\mathrm{B})}+1\right).
	\label{eq:widths}
	\end{equation}
~\\
Here $c_m=14.4$ is related to a magnetic cutoff, which is an additional parameter of the DQPM, and $C_q = \dfrac{N_c^2 - 1}{2 N_c} = 4/3$ and $C_g = N_c = 3$ are the {QCD} color factors for quarks and for gluons, respectively. We also assume that all (anti-)quarks have the same width. With the choice of Eq. (\ref{eq:spectral_function}), the complex self-energies for gluons $\Pi = M_g^2-2i \omega \gamma_g$ and for (anti)quarks $\Sigma_{q} = M_{q}^2 - 2 i \omega \gamma_{q}$ are fully defined via Eqs. (\ref{eq:Mg}), (\ref{eq:Mq}), (\ref{eq:widths}).

The strength of the interaction is defined by the coupling constant (squared) $g^2 = 4\pi\alpha_s$. This is an essential quantity which enters the definition of the DQPM thermal masses and widths and so determines all the microscopic properties as well as transport coefficients. In the DQPM this quantity can be extracted from lQCD in the following way. At vanishing chemical potential $g^2$ is obtained by a parametrization, introduced in Ref. \cite{Berrehrah:2016vzw}, of the lQCD results for the entropy density $s(T,\mu_B = 0)$ obtained by the BW Collaboration \cite{Borsanyi:2012cr,BORSANYI201499}
\begin{equation}
  g^2(T,\mu_B=0) = d \left( (s(T,0)/s^{QCD}_{SB})^e - 1 \right)^f.
  \label{eq:g2mub0}
\end{equation}
Here $s^{QCD}_{SB} = 19/9 \pi^2T^3$ is the Stefan-Boltzmann entropy density and $d = 169.934, e = -0.178434$ and $f = 1.14631$ are the dimensionless parameters. 

\begin{figure}[h!]
  \centering
  \includegraphics[width=\columnwidth]{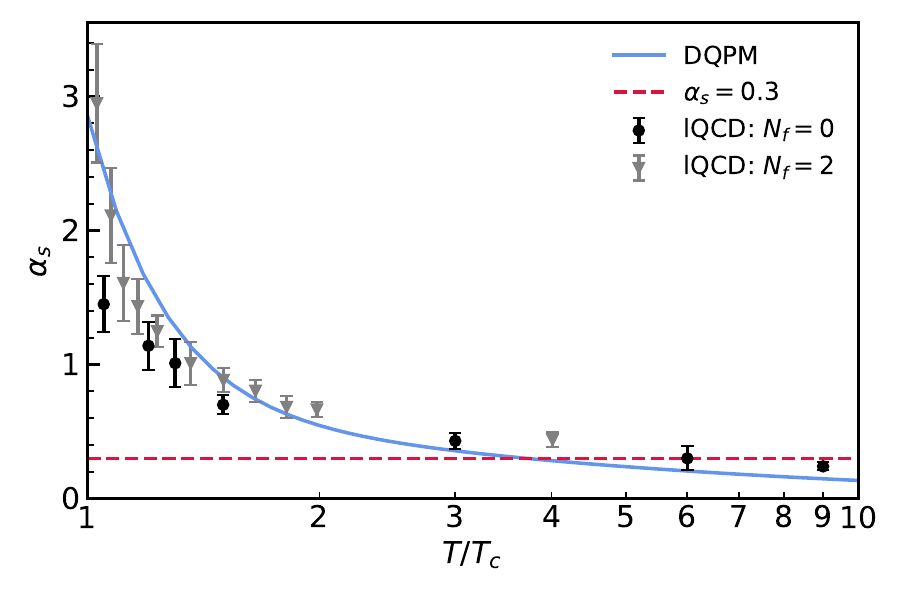}
  \caption{Running coupling constant $\alpha_s$ as a function of $T/T_c$. Blue solid line corresponds to the DQPM running coupling defined by Eq. \eqref{eq:coupling}, red dashed line indicates a constant value of 0.3. The lattice results for quenched QCD, $N_f = 0$, (black circles) are taken from Ref. \cite{Kaczmarek:2004gv} and for $N_f = 2$ (gray triangles) are taken from Ref. \cite{Kaczmarek:2005PRD}.
  }
  \label{fig:qhat_alphas}
\end{figure}

The running coupling constant $\alpha_s$ for $\mu_B=0$, defined by the Eq. \eqref{eq:g2mub0}, is shown in Fig. \ref{fig:qhat_alphas} as a blue solid line as a function of $T/T_c$. The lattice results for quenched QCD, $N_f = 0$, (black circles) are taken from Ref. \cite{Kaczmarek:2004gv} and for $N_f = 2$ (gray triangles) are taken from Ref. \cite{Kaczmarek:2005PRD}. The DQPM $\alpha_s$ rises in vicinity of $T_C$ in line with the lQCD data. The red dashed lines indicate a constant value of $\alpha_s =0.3$ often used in the pQCD models, in particular in the BAMPS calculations used for the model comparison which will be discussed in Section IV. 

The extension of coupling constant to finite baryon chemical potential, $\mu_B$, is performed by using a scaling hypothesis \cite{Cassing:2008nn}, which works up to $\mu_B \approx 450$ MeV, and which assumes that $g^2$ is a function of the ratio of the effective temperature $T^* = \sqrt{T^2 + \mu_q^2/\pi^2}$ and the $\mu_B$-dependent critical temperature $T_c(\mu_B)$ as
\begin{equation}
  \label{eq:coupling}
  g^2(T/T_c,\mu_B) = g^2\left(\frac{T^*}{T_c(\mu_B)},\mu_B =0 \right)
\end{equation}
with $T_c(\mu_B) = T_c \sqrt{1 - \alpha\mu_B^2}$, where $T_c$ is the critical temperature at vanishing chemical potential ($\approx$~0.158~GeV) and $\alpha \simeq 0.974$ GeV$^{-2}$.

Having the quasiparticle properties, dressed propagators and coupling constant as given by the DQPM, one can calculate the scattering cross sections as well as the transport coefficients of quarks and gluons in the QGP as a function of the temperature and the chemical potential \cite{Berrehrah:2013mua,Moreau:2019vhw}. The matrix elements $|M|^2$ of the corresponding processes are calculated in leading order.


\section{Transport coefficients in DQPM}
\label{sec:tcoef}

\subsection{On-shell case}

In the 'on-shell' case all energies of the partons are taken to be $E^2=\vec{p}^2+M^2$, where $M$ is the pole mass of quark or gluon spectral function which depends on $(T,\mu_B$) and defined by Eqs. (\ref{eq:Mg}),(\ref{eq:Mq}). The general expression for a transport coefficient in kinetic theory \cite{PhysRevD.37.2484, PhysRevC.57.889, Moore:2004tg,Berrehrah:2014kba} has the following form:
\begin{align}
  \label{eq:O_on}
  \langle \calO \rangle^{\text{on}} = \frac{1}{2E_i}\sum_{j=q,\bar{q},g}
   \int\frac{d^3p_j}{\dpi^3 2E_j} d_j f_j \int\frac{d^3p_1}{\dpi^3 2E_1}
  \nonumber\\
  \times \int\frac{d^3p_2}{\dpi^3 2E_2} (1 \pm f_1)(1 \pm f_2)\ \calO \ \Msq\
  \nonumber\\
  \times (2\pi)^4 \delta^{(4)}(p_i + p_j - p_1 - p_2)
  \nonumber\\
  =\sum_{j=q,\bar{q},g}\int\frac{d^3p_j}{\dpi^3} d_j f_j
   v_{\text{rel}} \int d\sigma^{\text{on}}_{ij \to 12}\ \calO \ (1\pm f_1) (1\pm f_2),
\end{align}
where $d_j$ is the degeneracy factor for spin and color ($2N_c$ for quarks and $2(N_c^2-1)$ for gluons); $f_j = f_j(E_j,T,\mu_q)$ are the Fermi distribution functions for quarks and $f_j = f_j(E_j,T)$ are the Bose distribution functions for gluons. Here $v_{\text{rel}}=F/(4E_iE_j)$ with $F$ being the flux of incident particles. The Pauli-blocking (-) and Bose-enhancement (+) factors account for the available density of final states. 

Depending on the choice of $\calO$ in equation \eqref{eq:O_on} one can refer to different transport coefficients:
\begin{itemize}
    \item $\calO = (p_T^{2} - p_T^{\prime 2})$ -- to the jet transport coefficient $\qhat$,
    \item $\calO = (E-E')$ -- to the energy loss $\Delta E = dE/dx$,
    \item $\calO = (p_L-p_L')$ -- to the drag coefficient $\mathcal{A}$,
    \item $\calO = 1$ -- to the scattering rate $\Gamma$, 
\end{itemize}
where $p_T$ and $p_L$ denote transverse and longitudinal momenta respectively.

We note that the transport coefficients for the charm quark have been studied within the DQPM in Refs. \cite{Berrehrah:2016vzw,Xu:2018gux,Cao:2018ews,Song:2019cqz}, and the scattering rate $\Gamma$ has been calculated in Ref. \cite{Moreau:2019vhw}. All transport coefficients generally depend on jet momentum, and medium properties, which can be characterize by their $T$ and $\mu_B$ dependencies.

The differential cross sections for the $ij\to 12$ scattering in Eq. \eqref{eq:O_on} is defined as
\begin{align}
  &d\sigma^{\text{on}}_{ij \to 12} = \frac{d^3p_1}{\dpi^3 2E_1} \frac{d^3p_2}{\dpi^3 2E_2}\ \nonumber\\
  & \times \dpi^4 \delta^{(4)}(p_i + p_j - p_1 - p_2) \frac{\Msq}{F}.
\end{align}

To simplify Eq. \eqref{eq:O_on}, we evaluate it in the CM frame. In this case we have
\begin{align}
  d\sigma^{\text{on}}_{\text{CM}} &= \frac{p_f\ d \Omega}{16 \pi^2 \sqrt{s}} 
  \frac{\Msq}{F^{\text{CM}}} = \frac{d \Omega}{64 \pi^2 s} \frac{p_f}{p_i} \Msq,
  \nonumber\\
  F^{\text{CM}} &= 4p_i^{\text{CM}}\sqrt{s},
\end{align}
and the final expression for the transport coefficient reads
\begin{align}
  \langle \calO \rangle^{\text{on}} &= 
  \sum_{j=q,\bar{q},g}\int\frac{d^3p_j}{\dpi^3} d_j f_j v_{\text{rel}}
  \label{eq:O_on1}\\ 
  &\times \int \frac{d \Omega}{64 \pi^2 s} \frac{p_f^{\text{CM}}}{p_i^{\text{CM}}}
  \Msq\ \calO\ (1 \pm f_1)(1 \pm f_2). \nonumber
\end{align}
Here the relative velocity in the CM frame is given by
\begin{align}
  v_{\text{rel}} =\frac{\sqrt{(p_i \cdot p_j)^2-m_i^2 m_j^2}}{4E_i E_j}=\frac{p_{CM}\sqrt{s}}{E_i E_j}.
  \label{eq:v_rel_cm}
\end{align}
$p_{CM}$ is the momenta of the initial $(i,j)$ as well as of the final quarks $(1,2)$ in the CM frame given by
\begin{align}
  p_{CM}=\frac{\sqrt{(s-(m_{i,1}-m_{j,2})^2)(s-(m_{i,1}+m_{j,2})^2)}}{2\sqrt{s}}.
  \label{eq:p_cm}
\end{align}


\subsection{Off-shell case}

In the off-shell case the energy of the partons and their momenta, are independent variables which follow off-shell dispersion relations: $\tilde{E}_{j}^2({\bf p})={\bf p}^2+M_{j}^{2}-\gamma_{j}^{2}$ and masses are distributed according to the spectral functions of Eq.~(\ref{eq:spectral_function}). To take this into account, we should replace the Lorentz-invariant on-shell phase space measure by the off-shell equivalent:
\begin{align}
  \frac{1}{2E} &\to \int\frac{d\omega}{\dpi}\ \tilde{\rho}(\omega,\vec{p})\ \theta(\omega);
  \nonumber\\
  \int\frac{d^3p}{\dpi^3 2E} &\to \int\frac{d^4p}{\dpi^4}\ \tilde{\rho}(\omega,\vec{p})\ \theta(\omega);
\end{align}
where $\tilde{\rho}(\omega,\vec{p})$ -- the \textit{renormalized} time-like spectral function which allows to keep the probabilistic interpretation for the $\tilde{\rho}(\omega,\vec{p})$ for the description of the scattering of real, i.e. time-like, partons here \cite{Moreau:2019vhw}:
\begin{equation}
  \tilde{\rho}(\omega,\vec{p}) = \frac{\rho(\omega,\vec{p}) \theta(p^2)}
  {\int_0^\infty \frac{d\omega}{2\pi} 2\omega \rho(\omega,\vec{p}) \theta(p^2)}.
\end{equation}

However, since the jet parton is ultrarelativistic and is not in the equilibrium with the QGP partons, it is still considered as an on-shell particle, the off-shell phase-space measure is imposed only for the medium partons. In this case, the transport coefficient $\calO$ reads
\begin{align}
  \langle \calO \rangle^{\text{off}} = \frac{1}{2E_i}\sum_{j=q,\bar{q},g}
  \int\frac{d^4p_j}{\dpi^4} d_j f_j \tilde{\rho}(\omega_j,\vec{p}_j) \theta(\omega_j)
  \nonumber\\
  \times \int\frac{d^3p_1}{\dpi^3 2E_1} \int\frac{d^4p_2}{\dpi^4} \tilde{\rho}(\omega_2,\vec{p}_2)\theta(\omega_2) 
  \nonumber\\
  \times (1 \pm f_1)(1 \pm f_2)\ \calO \ 
  \Msq\ (2\pi)^4 \delta^{(4)}(p_i + p_j - p_1 - p_2).
  \label{eq:O_off}
\end{align}

It is important to note that in the evaluations of the matrix elements we employ effective propagators as in the previous works \cite{Moreau:2019vhw, Soloveva:2019xph}. In case of massive quarks and gluons with non-zero widths the DQPM propagators reads
\begin{align}
    S_F(q,M_q) = i \dfrac{\slashed{q}+ M_q}{ q^2- M_q^2+2i \gamma_q q_0},\\
    G_F^{\mu \nu}(q,M_g) = -i \dfrac{g_{\mu \nu}- q_{\mu}q_{\nu}/M_g^2}{q^2 - M_g^2+2i \gamma_g q_0}.
    \label{prop_new}
\end{align}
Here $q$ is the 4-momentum of the exchanged parton, $M_{q,g}, \gamma_{q,g}$ are the pole masses and widths of the quark/gluon given by Eqs. (\ref{eq:Mg}) - (\ref{eq:widths}).


\subsection{Contributing interaction channels}
\label{ssec:contrib}

\subsubsection{Quark jet}

\begin{figure*}[th!]
  \includegraphics[height=2.7cm]{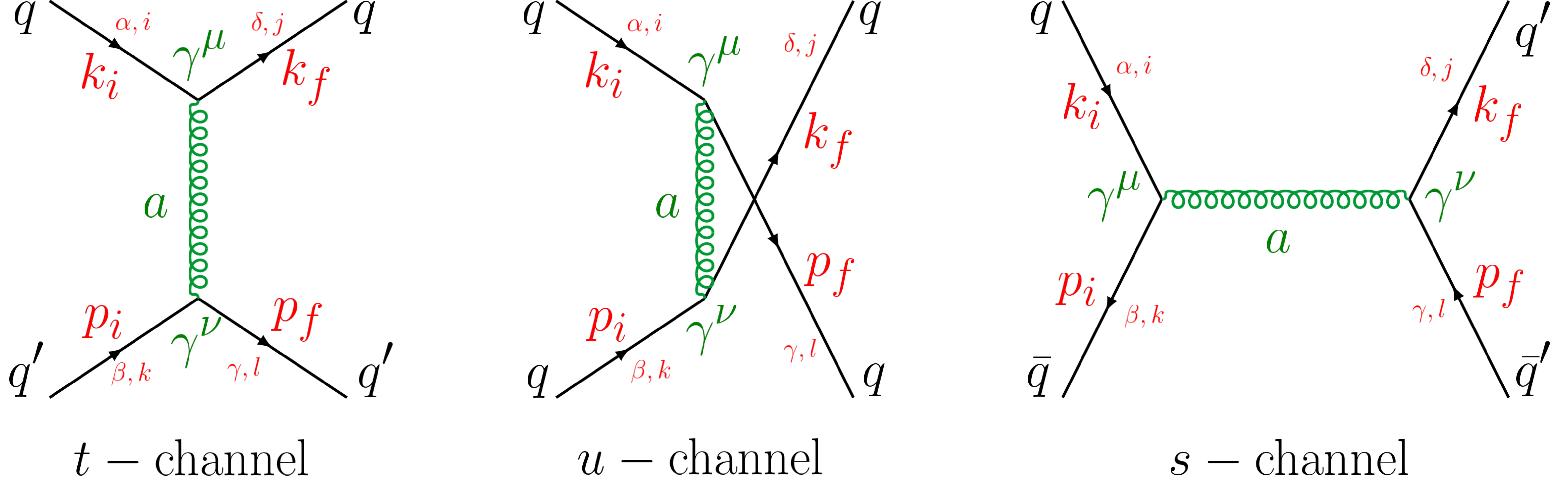}
  \includegraphics[height=2.7cm]{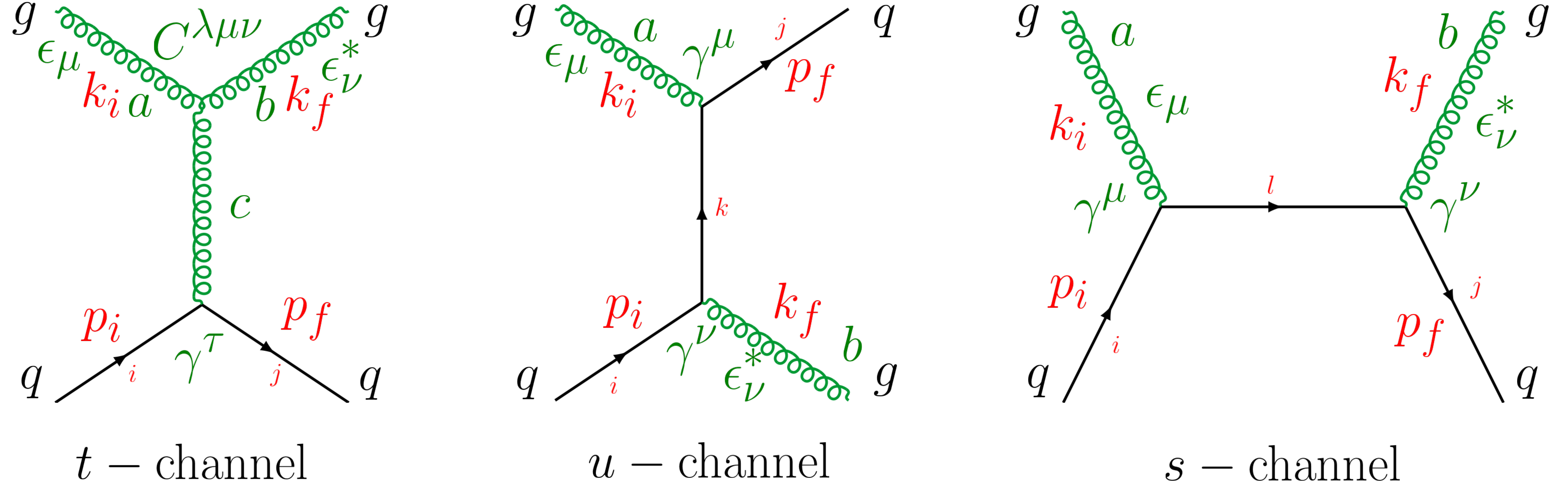}
  \includegraphics[height=2.7cm]{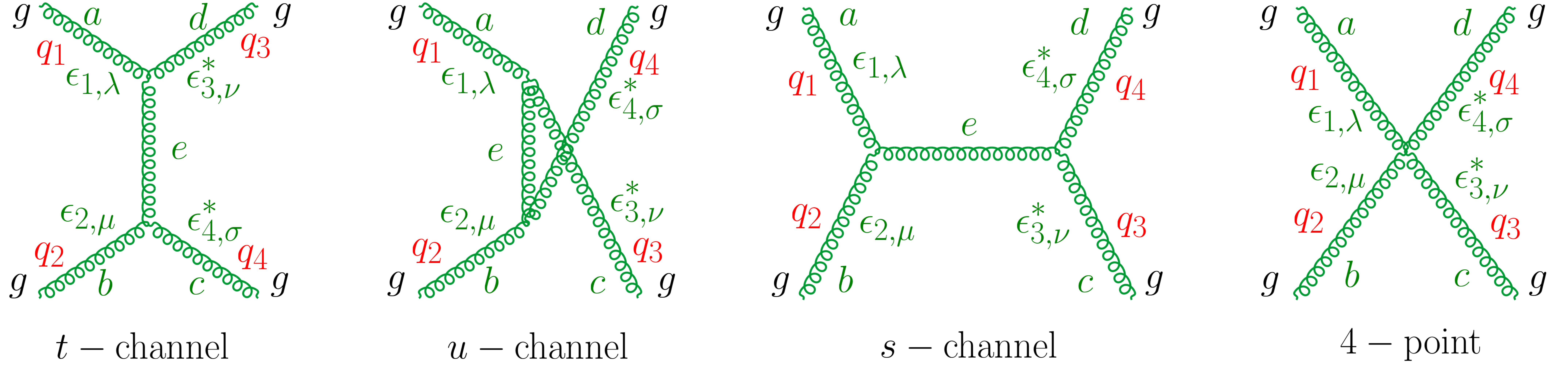}
  \caption{Leading-order Feynman diagrams for the elastic processes.}
  \label{fig:feynman}
\end{figure*}

In Fig. \ref{fig:feynman} we display the leading-order Feynman diagrams for the elastic processes which accounted in the calculations of jet transport coefficients.

The notation $\sum_{j=q,\bar{q},g}$ includes the contribution from all possible partons which in our case are the gluons and the (anti-)quarks of three different flavors ($u,d,s$). If we consider the jet as a $u-$quark, we have the following interaction channels:
\begin{itemize}
  \item $ uu \to uu$ ($t + u$ channels)
  \item $ u\bar{u} \to u\bar{u}$ ($t + s$ channels)
  \item $ ud \to ud$ ($t$ channel)
  \item $ u\bar{d} \to u\bar{d}$ ($t$ channel)
  \item $ us \to us$ ($t$ channel)
  \item $ u\bar{s} \to u\bar{s}$ ($t$ channel)
  \item $ ug \to ug$ ($t + u + s$ channels).
\end{itemize}

For $\mu_q=0$ there is no numerical difference between $ ud \to ud$/$ u\bar{d} \to u\bar{d}$ and $ us \to us$/$ u\bar{s} \to u\bar{s}$ processes (for $ uu \to uu$ this is not the case because different channels are included); considering this, the final summation reads
\begin{align}
  \langle \calO \rangle &= \langle \calO \rangle_{uu \to uu} + \langle \calO \rangle_{u\bar{u} \to u\bar{u}}
  + 2\langle \calO \rangle_{ud \to ud} 
  \nonumber\\
  &+ 2\langle \calO \rangle_{us \to us} + \langle \calO \rangle_{ug \to ug}.
\end{align}


\subsubsection{Gluon jet}

For the \textit{gluon}-jet the channels are as following:
\begin{itemize}
  \item $ gu \to gu$ ($t + u + s$ channels)
  \item $ g\bar{u} \to g\bar{u}$ ($t + u + s$ channels)
  \item $ gd \to gd$ ($t + u + s$ channels)
  \item $ g\bar{d} \to g\bar{d}$ ($t + u + s$ channels)
  \item $ gs \to gs$ ($t + u + s$ channels)
  \item $ g\bar{s} \to g\bar{s}$ ($t + u + s$ channels)
  \item $ gg \to gg$ ($t + u + s$ channels + 4 point amplitude)
\end{itemize}

Summing up all the contributions, we get for the \textit{gluon} jet:
\begin{equation}
  \langle \calO \rangle=4\langle \calO \rangle_{gu \to gu} + 2\langle \calO \rangle_{gs \to gs}
  + \langle \calO \rangle_{gg \to gg}.
\end{equation}


\section{Results }
\label{sec:results}

\subsection{\texorpdfstring{$\qhat$}{qhat} coefficient}

In this section we show the DQPM results on the temperature, energy, $\mu_B$ dependencies of the $\qhat$ coefficient and compare the DQPM results with other model calculations. The on-shell and full off-shell DQPM results, presented here, are performed according to Eqs. (\ref{eq:O_on}) and (\ref{eq:O_off}), respectively, using $(T,\mu_B)-$dependent $\alpha_S$ (Eqs.(\ref{eq:g2mub0}),(\ref{eq:coupling})) in both vertices of leading order Feynman diagrams (if not specified explicitly for the so called 'model study').


\subsubsection{On-shell vs off-shell dependence}

When an energetic jet parton penetrates the sQGP, it scatters with thermalized partons from the QGP medium and is loosing its energy and momentum. Since in the DQPM the QGP partons are strongly interacting dynamical quasiparticles, it is important to study the role of off-shell effects on $\qhat$. For the purpose we compare transport coefficients for elastic scattering of the jet partons (which follows an on-shell dispersion relation due to its high energy) with medium on-shell and off-shell partons. We note that the off-shellness reduces cross sections, the interaction rate and transport coefficients \cite{Moreau:2019vhw} due to the fact that the low mass part of spectral functions contributes with a larger weight according to the statistical occupation probability defined by the Fermi or Bose distributions. 

\begin{figure}[h!]
  \centering
  \includegraphics[width=\columnwidth]{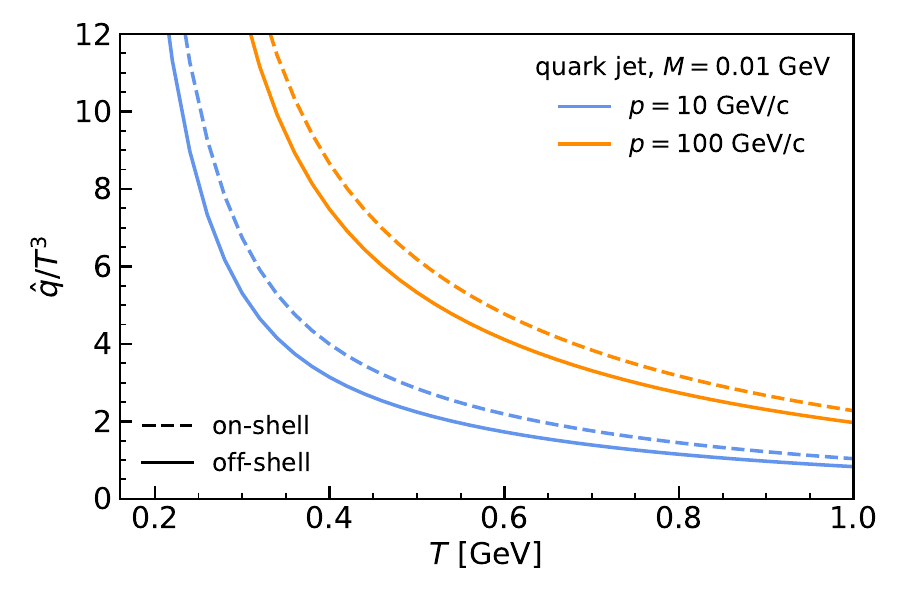}
  \caption{Comparison between the temperature dependent scaled off-shell (solid lines) and on-shell (dashed lines) $\qhat$ coefficients for a quark jet with 10 GeV (blue (lower) lines) and 100 GeV (orange (upper) lines) momentum.}
  \label{fig:qhat-T_OnOff}
\end{figure}

In Fig. \ref{fig:qhat-T_OnOff} we show the comparison between on-shell and off-shell results from Eq. \eqref{eq:O_on} and Eq. \eqref{eq:O_off} for $\hat{q}/T^3$ of a jet quark whose mass is $0.01$ GeV and momentum 10 and 100 GeV/c. For a jet quark with both 10 GeV/c and 100 GeV/c momentum the on-shell case gives larger results over the entire temperature range than the off-shell case. The off-shell effect is stronger for low $T$ where the elastic scattering (addressed in the present study) is expected to play a dominant role compared to radiative processes, while with increasing $T$ the difference between on-shell and off-shell results is getting small.


\subsubsection{Dependence on the mass of jet quark}

\begin{figure}[th!]
  \centering
  \includegraphics[width=\columnwidth]{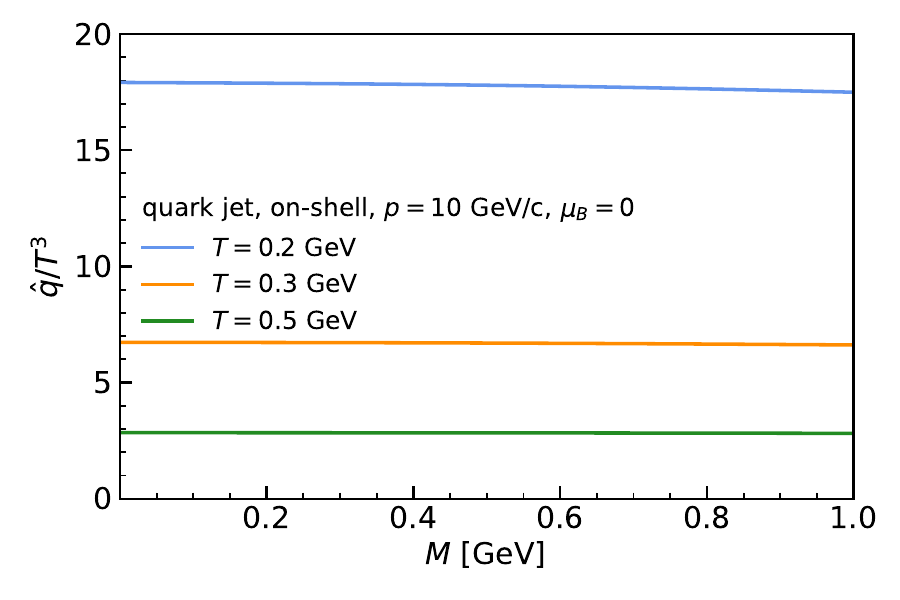}
  \includegraphics[width=\columnwidth]{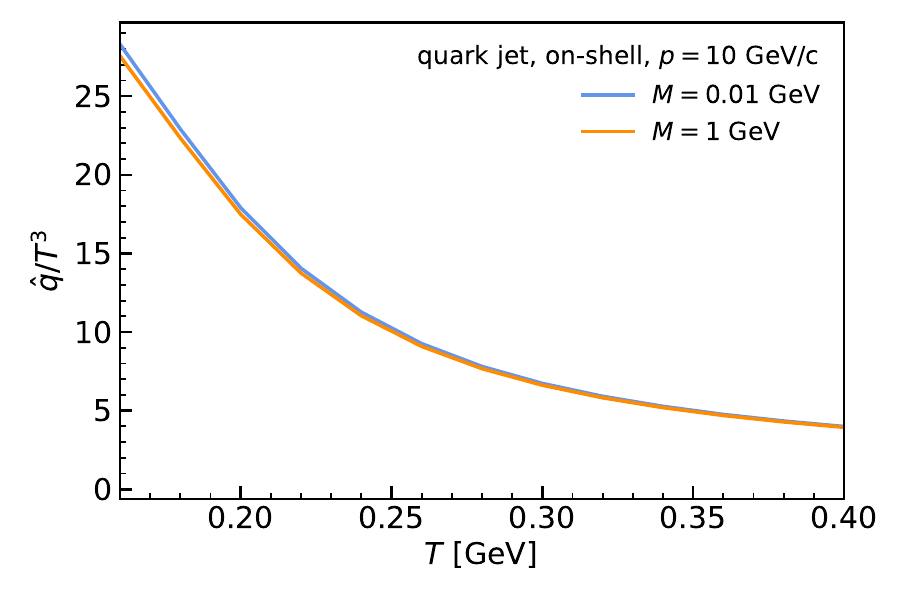}
  \caption{Upper plot: mass dependence of the scaled $\qhat/T^3$ coefficient for a quark jet with momentum $p=10$ GeV/c at $\mu_B=0$ and for different medium temperatures $T=0.2$ GeV (blue (upper) line), $T=0.3$ GeV (orange (middle) line) and $T=0.4$ GeV (green (lower) line). Lower plot: temperature dependence of the scaled $\qhat/T^3$ coefficient for a quark jet with momentum $p=10$ GeV/c at $\mu_B=0$ for different jet masses $M=0.01$ GeV (blue (upper) line) and $M=1$ GeV (orange (lower) line).
  }
  \label{fig:qhat_m}
\end{figure}

Now we investigate the dependence of the $\qhat$ coefficient on the 'virtuality', i.e. the mass of the jet parton. Fig. \ref{fig:qhat_m} shows how the $\qhat/T^3$ coefficient changes with $M$ and $T$. The upper part of Fig. \ref{fig:qhat_m} shows the mass dependence of the scaled $\qhat/T^3$ coefficient for a quark jet with momentum $p=10$ GeV/c at $\mu_B=0$ and for different temperatures $T=0.2$ GeV (blue line), $T=0.3$ GeV (orange line) and $T=0.4$ GeV (green line). As follows from the plot, $\qhat/T^3$ shows a very weak dependence on the mass $M$ at $T=0.2$~GeV and getting flat with increasing $T$. The lower part of Fig. \ref{fig:qhat_m} illustrates the temperature dependence of the scaled $\qhat$ coefficient for a quark jet with momentum $p=10$ GeV/c at $\mu_B=0$ for different jet masses $M=0.01$ GeV (blue line) and $M=1$ GeV (orange line). As one can see, the $\qhat$ coefficient remains almost unchanged for masses from 10 MeV to 1 GeV. The reason for this behavior is due to the fact that $\qhat$ depends in general on the jet energy $E=\sqrt{\vec{p}^2+M^2}$, which is constructed from the jet mass and jet momentum. For momenta $p > 10$ GeV the jet mass is practically negligible even for $M = 1$ GeV. In principle, the virtuality can be as large as jet energy. However, the virtuality survives only for a time of inverse virtuality, which is normally much shorter than the thermalization time of QGP in heavy-ion collisions. For example, the virtuality of 1 GeV survives 0.2 fm/c.

We note that for all further results we will use $M=0.01$ GeV (if not stated explicitly) which corresponds to the jet parton which lost its virtuality when entering the medium in heavy-ion experiments \cite{Zapp:2008gi}.


\subsubsection{Channel decomposition}

\begin{figure*}[th!]
  \centering
  \includegraphics[width=0.95\textwidth]{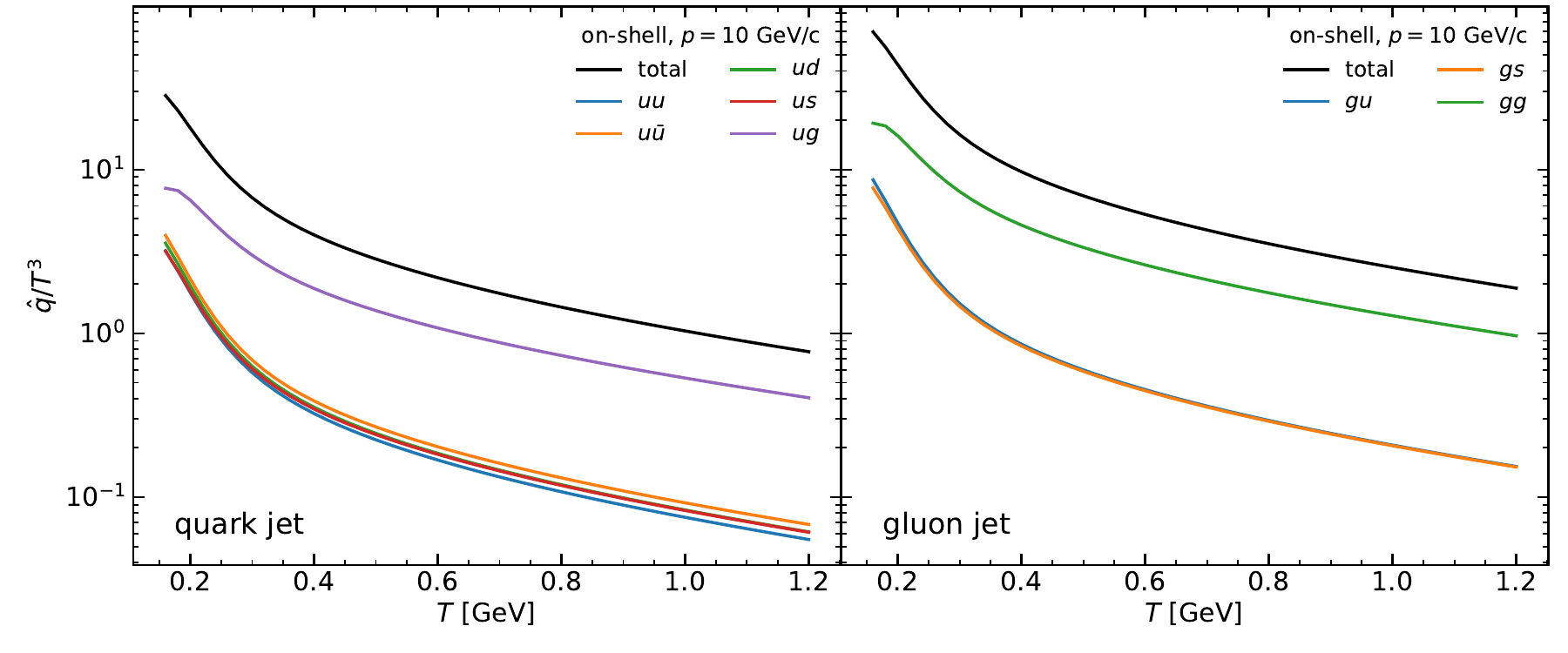}
  \caption{Channel decomposition of the temperature dependent scaled on-shell $\qhat$ coefficient for a $u$-quark (left) and gluon (right) jet. Black (upper) lines correspond to the total value of $\qhat/T^3$. The color lines correspond to the contribution of individual channels, i.e. left plot: $uu$ (blue (6-th from top)), $ud$ (green (4-th from top)), $us$ (red (5-th from top)), $u\bar u$ (orange (3rd from top)), $ug$ (violet (2nd from top));
  right plot: $gu$ (blue (3rd from top)), $gs$ (orange (4-th from top)), $gg$ (green (2nd from top)).}
  \label{fig:qhat_chan_dec}
\end{figure*}

As described in Sec. \ref{ssec:contrib}, a final value of the $\qhat$ coefficient is made up of several contributions from the different interaction processes. Therefore, it is interesting to investigate how each interaction channel contributes to the final value of the jet transport coefficient. In Fig. \ref{fig:qhat_chan_dec} we show these contributions to the temperature dependent $\qhat$ coefficient for a quark (left) and gluon (right) jet in the on-shell case.

As one can see from the figure, the highest contribution for both quark and gluon jets comes from the interaction with the medium gluons. The reason for this comes from a large number of possible interaction channels between quarks with gluons or gluons with gluons.

For a quark jet a very small difference shows up between the $qq$ interaction and $q\bar{q}$ contributions, in spite of that $uu \to uu$ and $u\bar{u} \to u\bar{u}$ differs by their interaction channels. We note that for the $uu \to uu$ the forward $u$-quark is taken as the jet $u$-quark when transformed from center-of-mass system to the rest frame of the medium (similar to the BAMPS calculations in Ref. \cite{Senzel:2020tgf}). For the gluon jet a $gq$ interaction and a $g\bar{q}$ interaction give identical contributions since the gluon does not distinguish between a quark and an antiquark and all interaction channels are the same. We note that the $\hat q$ of a gluon jet differs from the one of a quark jet by about a color factor $9/4$.

\subsubsection{Chemical potential dependence}

\begin{figure}[ht!]
  \centering
  \includegraphics[width=\columnwidth]{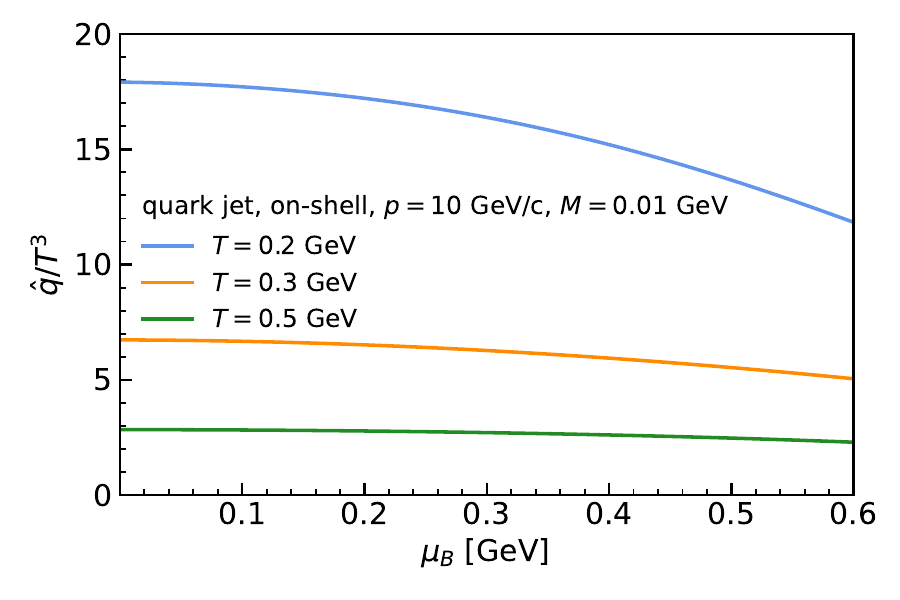}
  \includegraphics[width=\columnwidth]{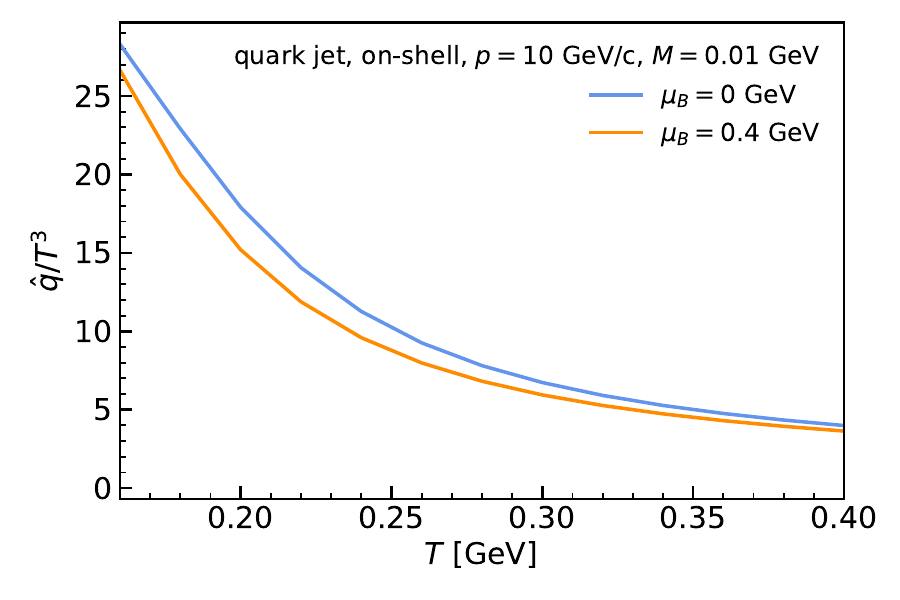}
  \caption{Upper plot: Chemical potential $\mu_B$ dependence of the scaled $\qhat/T^3$ coefficient for a quark jet with momentum $p=10$ GeV/c at $M=0.01$ GeV and for different medium temperatures $T=0.2$ GeV (blue (upper) line), $T=0.3$ GeV (orange (middle) line) and $T=0.4$ GeV (green (lower) line). Lower plot: temperature dependence of the scaled $\qhat/T^3$ coefficient for a quark jet with momentum $p=10$ GeV/c and jet mass $M=0.01$ GeV for different chemical potentials $\mu_B=0$ GeV (blue (upper) line) and $\mu_B=0.4$ GeV (orange (lower) line).
  }
  \label{fig:qhat_mub}
\end{figure}

The DQPM is able to describe not only the temperature dependence of microscopic properties, but also the dependence on baryon chemical potential, since $\mu_B$ enters explicitly the equations for the parton masses, widths, and coupling constant. It has been shown in Ref. \cite{Moreau:2019vhw} that all mentioned quantities as well as differential and total cross sections decrease with the increase of the baryon chemical potential. Based on this fact, we assume that the $\qhat$ coefficient should also decrease for a finite $\mu_B$. To verify our assumption, we have calculated the $\qhat$ coefficient for $\mu_B = 0.5$ GeV and compare it with the calculations for $\mu_B=0$. The results are presented in Fig. \ref{fig:qhat_mub}. The upper part of Fig. \ref{fig:qhat_mub} shows the chemical potential $\mu_B$ dependence of the scaled $\qhat/T^3$ coefficient for a quark jet with momentum $p=10$ GeV/c at $M=0.01$ GeV and for different medium temperatures $T=0.2$ GeV (blue line), $T=0.3$ GeV (orange line) and $T=0.4$ GeV (green line). The lower part presents the temperature dependence of the scaled $\qhat/T^3$ coefficient for a quark jet with momentum $p=10$ GeV/c and jet mass $M=0.01$ GeV for different chemical potentials $\mu_B=0$ GeV (blue line) and $\mu_B=0.4$ GeV (orange line). Consequently, the $\qhat/T^3$ coefficient decreases with increasing baryon chemical potential. This decrease is stronger for $T=0.2$ GeV than for high $T$.

We note that the dominant effect on the $\mu_B$ dependence of $\qhat$ comes from the behavior of the coupling constant. In general, there are 2 quantities in the equation for the transport coefficients which strongly depend on $\mu_B$: the coupling constant $\alpha_S(T,\mu_B)$ and the distribution function $f(E,T,\mu_B)$ for quarks. There is also a $\mu_B$ dependence for the masses/widths of the medium/intermediate(exchange) quarks/gluons, but it is relatively small. Even though the distribution function increases with increasing $\mu_B$, the effect of decreasing $\alpha_S(T,\mu_B)$ is stronger; so eventually we see a reduction of the $\qhat/T^3$ coefficient.

Such a $\mu_B$ dependence of $\qhat$ might show up in heavy-ion collisions at SPS and low RHIC energies where the created QGP matter has a non-zero baryon chemical potential.


\subsubsection{\texorpdfstring{$\qhat$}{qhat} coefficient vs \texorpdfstring{$T$}{T}: model comparison}

\begin{figure}[t!]
  \centering
  \includegraphics[width=0.48\textwidth]{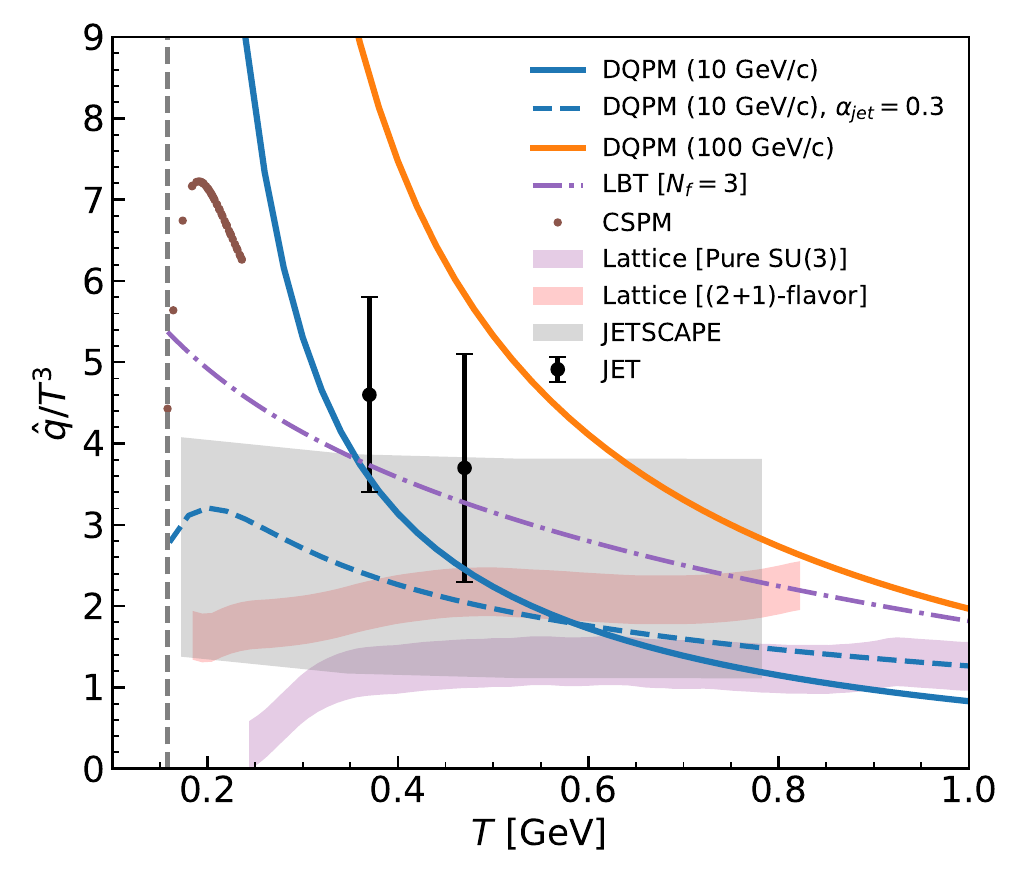}
  \caption{Temperature dependence of the scaled jet transport coefficient $\hat{q}/T^3$. The off-shell DQPM results are represented for a quark jet with mass $M=0.01$~GeV and momentum 10 GeV/c (blue (lower) line) and 100 GeV/c (orange (upper) line). The blue dashed line shows the DQPM result with $\alpha_{S}^{jet}=0.3$ at the jet parton vertices. The purple dashdotted line represents the LBT results for $N_f=3$ and $p=10$ GeV/c \cite{He:2015pra}, while the red (upper) and purple (lower) areas represent lQCD estimates \cite{kumar2020jet} for pure SU(3) gauge theory and (2+1) flavour QCD, respectively, in the limit of an infinitely hard jet parton. The gray area corresponds to the results from the JETSCAPE Collaboration ($p = 100$ GeV/c) \cite{JETSCAPE:2021ehl}. The black dots represent the phenomenological extraction by the JET Collaboration presented for $p=10$ GeV/c \cite{Burke:2013yra}, while the brown dots show the results from the color string percolation model (CSPM) \cite{Mishra:2021yer}. The vertical gray dashed line indicates the critical temperature $T_C=0.158$ GeV.
  }
  \label{fig:qhat-T}
\end{figure}

In Fig. \ref{fig:qhat-T} we show the temperature dependence of the scaled $\qhat/T^3$ transport coefficient for $\mu_B=0$ for different approaches: 

{\bf 1)} The solid blue and orange lines show the main DQPM results for $\qhat/T^3$ obtained by full off-shell calculations -- according to Eq. (\ref{eq:O_off}) using $(T,\mu_B)-$dependent $\alpha_S$ (Eqs.(\ref{eq:g2mub0}),(\ref{eq:coupling})) in both vertices of leading order Feynman diagrams -- for a quark jet with momentum $p = 10$ GeV/c and $p = 100$ GeV/c, respectively. One can see a strong increase of $\qhat/T^3$ when approaching to $T_c$.

{\bf 2)} We compare the DQPM results with the recent lattice calculations \cite{kumar2020jet} for the SU(3) pure glue plasma using the Wilson gauge action (purple area) and (2+1) flavor QCD (red area) calculations. As stated in Ref. \cite{kumar2020jet} these lattice calculations include a single parton undergoing a single scattering of the medium simulated on the lattice, the momentum dependence of the jet parton is not accounted for and $\qhat/T^3$ is considered in the limit of an infinitely hard jet parton which leads to a rather different behavior at lower $T$.

{\bf 3)} Furthermore, we show the results from a phenomenological extraction of $\qhat$ from heavy-ion data:\\
i) the solid black dots with error bars present the $\qhat/T^3$ estimated by the JET Collaboration \cite{Burke:2013yra} from five different models (GLV-CUJET, HT-M, HT-BW, MARTINI and MCGILL-AMY) for the parton energy loss of a jet quark of 10 GeV energy in a dense medium (simulated by CUJET) by constraining the model parameters from the experimental data on the suppression factors $R_{AA}$ of large transverse momentum hadron spectra in heavy-ion collisions at both RHIC and LHC energies.\\
ii) The gray area shows the result on $\qhat/T^3$ for a jet quark of $p=100$ GeV/c evaluated by the JETSCAPE \cite{JETSCAPE:2021ehl} simulation framework (which incorporates multi-stage jet evolution, using the Matter model \cite{Majumder:2013re} at high virtuality scale, and the LBT model \cite{Cao:2017hhk} at low virtuality scale and relativistic hydrodynamics VISH2+1 for the QGP evolution \cite{Song:2007fn}), assuming the pQCD based temperature dependent form of $\qhat$ with few parameters, which have been obtained using a Bayesian analysis for a simultaneous calibration of jet quenching data on $R_{AA}$ at various centrality bins in Au+Au collisions at $\sqrt{s_{NN}}=$ 200 GeV, and in Pb+Pb collisions at $\sqrt{s_{NN}}=$ 2.76 TeV and 5.02 TeV. We note that the 'phenomenological' $\qhat$ - extracted from the experimental data - contains the information about all interactions of the jet parton with the medium, including the radiative processes, which notably contribute to the jet attenuation at high $p_T$ and explanation of the suppression factors $R_{AA}$ in pQCD models and which are not accounted for in our evaluations here within non-perturbative effective model DQPM.\\
iii) The brown dots show the results from the color string percolation model (CSPM) \cite{Mishra:2021yer}, based on the idea of strong string interactions leading to string cluster formation in the transverse plane by the percolation mechanism. The $\hat q$ has been evaluated using the percolation density parameters extracted by fitting transverse momentum spectra of hadrons from $p+p$ and $A+A$ collisions of different centralities at LHC energies. One can see that $\hat q/T^3$ decreases with increasing temperature due to a larger quenching for the hot system.

{\bf 4)} In order to see the influence of the non-perturbative effects on $\hat q$ we show the comparison of our DQPM results to the pQCD calculations. The results for elastic scattering of a jet quark with $p=10$ GeV/c obtained with the pQCD based LBT model for $N_f=3$ \cite{He:2015pra} are shown by purple dashed line. In the LBT model the LO-HTL approach for the Debye mass is employed for the regularization of the infrared divergence. In case of the LBT (and BAMPS) model one can estimate the high energy asymptotic behavior $E \gg T$ by employing a small-angle approximation for the differential cross-section. Following Ref. \cite{He:2015pra} the $\qhat$ coefficient is given by
\begin{equation}
  \hat{q}=\frac{T^3 C_F 42\zeta(3)}{ \pi} \alpha_s^2 \mathrm{ln} \left(\frac{q_{max}^2}{4\mu_D^2} \right),
  \label{eq:qhat-highElim}
\end{equation}
where $q_{max}$ is an UV cutoff, which can be chosen as $\sqrt{c E_{jet} T}$, with $E_{jet}$ being the jet energy, $c = 4- 6$: for the LBT model $c = 5.8$ (quark) \cite{He:2015pra}; with fixed $\alpha_S=0.3$ while $\zeta(3) \simeq 1.202$ is the Ap\'ery's constant. This asymptotic behavior coincides with the LO-HTL leading-log expression for $q_{max} \gg T$ \cite{Arnold:2008vd, Caron-Huot:2008zna}. 

As one can see from Fig. \ref{fig:qhat-T}, the DQPM and LBT (LO-HTL) results show a decrease of $\qhat$ with temperature. However, for temperatures close to $T_C$ we see a significant increase of $\qhat$ for the DQPM approach. Such a temperature dependence of $\qhat$ in the DQPM originates dominantly from the temperature dependence of the coupling constant $\alpha_s$ (Eq. \eqref{eq:g2mub0} depicted in Fig. \ref{fig:qhat_alphas}) while the LBT model assumes a constant $\alpha_S=0.3$. The DQPM results at large $T$ is lower than the pQCD LBT results due to a strong decrease of $\alpha_S(T)$ with temperature. 

In order to explore the influence of $\alpha_s(T,\mu_B)$ on $\qhat$ we perform a 'model study' for a qualitative comparison with pQCD results by replacing the $\alpha_s(T,\mu_B)$ at the 'hard' vertex of "jet parton - exchange parton - jet parton" by the constant $\alpha_{jet}\equiv \alpha_S=0.3$ (as used in the LBT model), while keeping the original $\alpha_s(T,\mu_B)$ at the 'thermal' vertex of "QGP parton - exchange parton - QGP parton". As it was shown in \cite{Carrington:2019ggz} for the QED case for the collective modes in the thermal plasma the natural renormalization scale is the temperature, therefore it is reasonable to consider $\alpha_s(T)$. This study could be also motivated by the fact that the jet parton is not in thermal equilibrium with the QGP medium. We note that an idea of using two different coupling constants in 'hard' and 'thermal' vertices has been explored early in Ref. \cite{Cao:2017hhk}. The result of our 'model study' is shown in Fig. \ref{fig:qhat-T} by the blue dashed line for a quark jet with momentum $p = 10$ GeV/c. One can see that such a replacement of $\alpha_S$ at the vertex of the jet parton by 0.3 leads to a significant reduction of $\qhat/T^3$ at low $T$. Moreover, the $T$-dependence of $\qhat$ becomes more similar to the pQCD model at large $T$. This illustrates a strong influence of the temperature dependent coupling constant on $\qhat$. 

We mention that in general, there are four effects that make the DQPM results different from the pure pQCD calculation:\\
- non-perturbative origin of the strong coupling $\alpha_s$ which depends on ($T,\mu_B$); \\
- finite masses of the intermediate parton propagators (analogy to the screening masses used for the regularization of the infrared divergence in the pQCD); \\
- finite widths in the intermediate parton propagators;\\
- finite masses and widths of medium partons.\\
In particular, \\
$\bullet$ Strong coupling is dominantly responsible for the sensitivity to the properties of the QCD medium since it enters the definitions of thermal masses/widths and scattering amplitudes. The strong temperature dependence of the coupling leads to the strong temperature dependence of the transport coefficients.\\
$\bullet$ Finite masses of the intermediate parton propagators play the same role as the Debye screening mass in the HTL calculations providing the cut-off effect and the general suppression for the differential cross sections.\\
$\bullet$ Finite masses of the medium partons have three effects on the total value of the transport coefficients. Firstly, they enter the expression for the scattering amplitude and have a large effect for small scattering energies. For high energies (which is the case for jet partons), however, the effect of finite masses becomes negligible. Secondly, parton masses enter the definition of transport coefficients, which leads to an increase of $\qhat$.
Thirdly, parton masses enter the distribution function of thermal partons $f(E,T,\mu_B)$ leading to a strong suppression of the transport coefficients, i.e. fewer scattering centers. This effect is dominant.\\
$\bullet$ The finite widths of partons also have a small effect on the scattering amplitudes, but are important for the off-shell calculations as they define the shape of the spectral function.\\
Thus, eventually a large sensitivity of the jet energy loss to the properties of the QCD medium comes not only from the strong coupling but from all aspects of the DQPM model.

As follows from Fig. \ref{fig:qhat-T} there are large model uncertainties in the determination of $\qhat$ from both theoretical and phenomenological side.


\subsubsection{Energy dependence of \texorpdfstring{$\qhat$}{qhat}}

\begin{figure*}[th!]
  \centering
  \includegraphics[width=\textwidth]{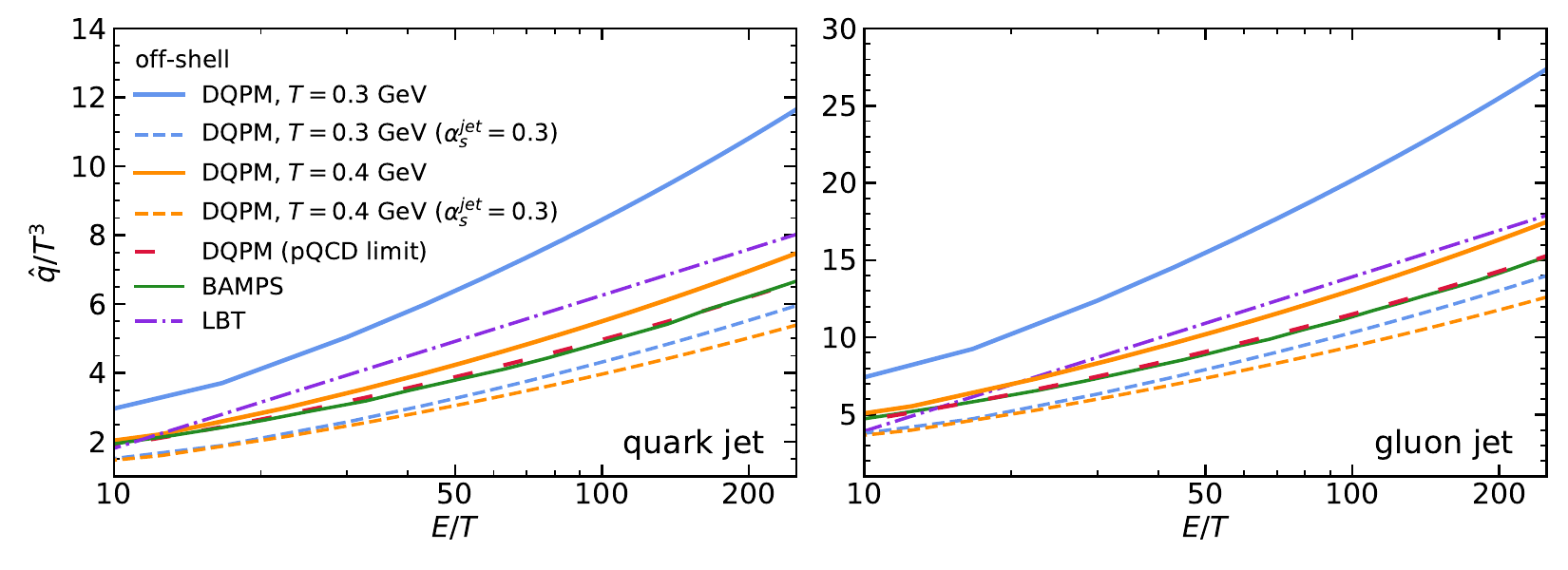}
  \caption{The scaled $\qhat/T^3$ coefficients as a function of $E/T$ for a quark jet (left) and a gluon jet (right) for different medium temperatures. The blue solid (upper) and orange solid (middle) lines represent the off-shell DQPM result for $T=0.3$ and $0.4$ GeV, respectively. The dashed lines of the same color represent DQPM results with $\alpha_{jet}=0.3$ at jet parton vertices for for $T=0.3$ and $0.4$ GeV, respectively. Red loosely dashed line represents the DQPM result in the pQCD limit. The green solid (lower) line represents BAMPS results \cite{Senzel:2020tgf} and the purple dashdotted line stands for the LBT model results \cite{He:2015pra}. All model calculations include only elastic energy loss.}
  \label{fig:qhat_p}
\end{figure*}

Figure \ref{fig:qhat_p} shows the scaled $\qhat/T^3$ coefficients for a quark (left) and a gluon (right) jet for elastic scattering with off-shell medium partons (from Eq. \eqref{eq:O_off}) as a function of the ratio of the jet energy over temperature $E/T$. The blue and orange lines represent the off-shell DQPM result for $T=0.3$ and $0.4$ GeV, respectively. The dashed lines of the same color represent DQPM results with $\alpha_{jet}=0.3$ at jet parton vertices for for $T=0.3$ and $0.4$ GeV respectively. Red dashed line represents the DQPM result in the pQCD limit. The green line represents BAMPS results \cite{Senzel:2020tgf} and the purple line stands for the LBT model results \cite{He:2015pra}. All model calculations include only elastic energy loss.

For all temperatures, the $\qhat/T^3$ coefficient logarithmically increases with the increasing momentum $p$ of the jet. This momentum dependence is in agreement with the asymptotic behavior of $\qhat$ for pQCD. 

In order to elucidate the systematic differences between the DQPM, LBT and BAMPS evaluations, we consider the pQCD limit for the DQPM (red dashed lines in Figure \ref{fig:qhat_p}):\\
$\bullet$ all partons are on-shell, i.e. widths are neglected;\\
$\bullet$ the exchange parton has a Debye mass in case of gluon $\mu_D^g= \dfrac{8\alpha_S}{\pi}(N_c+N_f)T^2$ or quark $M_D^q= \dfrac{2 \alpha_S}{\pi} C_F T^2$, while scattered partons are assumed to be massless; \\
$\bullet$ the DQPM coupling is fixed to $\alpha_S=0.3$ as in the LBT and BAMPS models;\\
$\bullet$ the classical (Maxwell-Boltzmann) statistics is employed.\\
In the high energy limit $E /T\gg 1$ with $\alpha_s = 0.3$, there is a logarithmic scaling of $\qhat/T^3 = const \ \mathrm{ln} \left(\frac{q_{max}^2}{4\mu_D^2} \right) = const \ \mathrm{ln} \left(\frac{E}{T} \right)$ as it follows from Eq. \eqref{eq:qhat-highElim}. A similar asymptotic behavior can be seen for the energy loss as well. Due to the different Debye masses for the LBT and BAMPS models, the results for $\qhat/T^3$ differ at high $E/T$. 

As seen from Fig. \ref{fig:qhat_p} for a quark jet (left plot), at low $T$ the DQPM shows a substantially larger $\qhat/T^3$ than the pQCD results of BAMPS and LBT models, due to the rise of $\alpha_S$ near $T_C$, while at high $T$ the DQPM results approach the pQCD predictions. However, when replacing $\alpha_S \to \alpha_{jet}=0.3$ at the jet parton vertices, the DQPM result decreases for low $T$ and becomes even smaller than the pQCD  models BAMPS and LBT. Moreover, the DQPM result in the pQCD limit (discussed above) is identical to the result of the BAMPS model (since the same Debye mass has been used in our calculations) for all $E/T$ models at low $T$, too. For the gluon jet (right plot) the DQPM shows again a reasonable agreement with pQCD models - BAMPS and LBT - for the $\alpha_{jet}=0.3$ and for the pQCD limit cases.

\subsection{Energy loss \texorpdfstring{$\Delta E$}{Δ E} coefficient}

\begin{figure*}[th!]
  \centering
  \includegraphics[width=\textwidth]{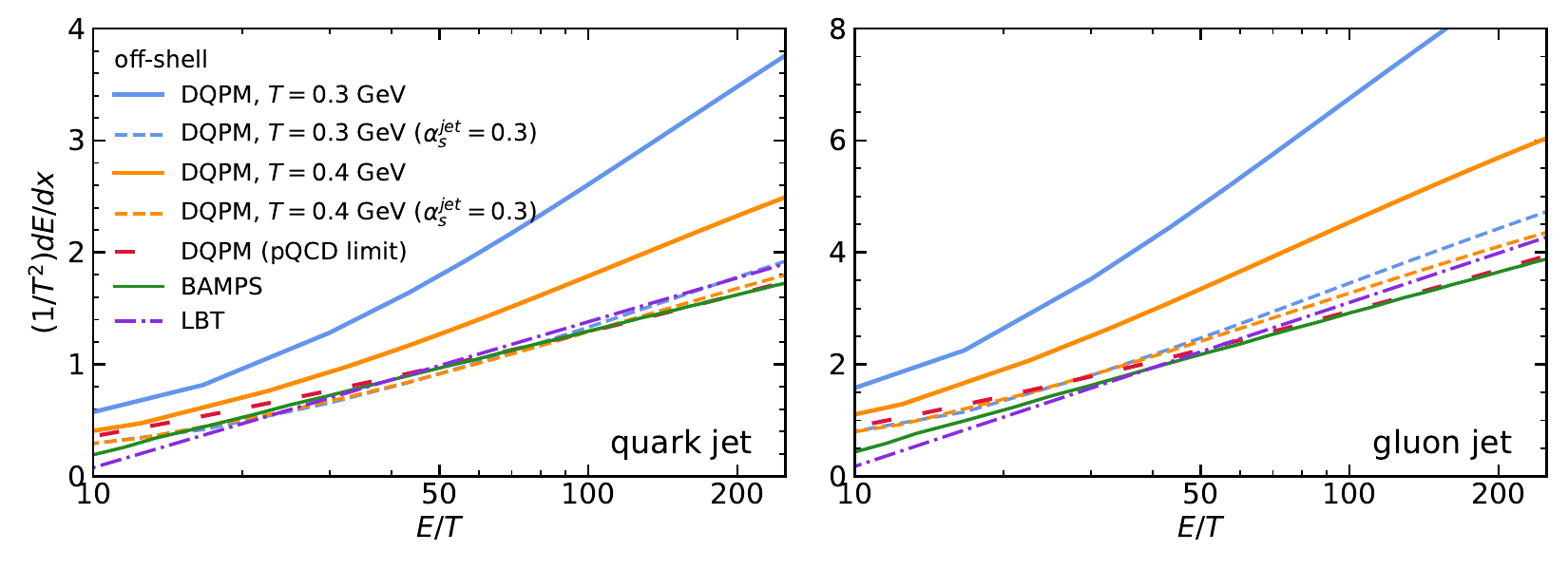}
  \caption{The scaled energy loss $(1/T^2) dE/dx$ as a function of $E/T$ for a quark jet (left) and gluon jet (right) with mass $M = 0.01$ GeV. The color coding of the lines for different models are the same as in Fig. \ref{fig:qhat_p}.}
  \label{fig:dE_p}
\end{figure*}

We proceed with the evaluation of the energy loss by the binary collisions. Fig. \ref{fig:dE_p} shows the scaled energy loss $1/T^2 \Delta E=1/T^2 dE/dx$ in the on-shell case for a quark jet with mass $M = 10$ MeV for medium temperatures $T=0.3$ and $0.4$ GeV.  As seen from Fig. \ref{fig:dE_p} $1/T^2 dE/dx$ shows similar to $\qhat/T^3$ a logarithmic increase with $E/T$. Similarly to the jet coefficient $\hat q$, one can consider the asymptotic behavior of energy loss within the small-angle approximation \cite{He:2015pra}:
\begin{equation}
  dE/dx=C_i \frac{3 \pi}{2} \alpha_s^2 T^2 \ln\left( \frac{q_{max}}{4 \mu_D^2} \right ).
\end{equation}
Here $C_i = C_F/ C_A$ denotes the color factor, $q_{max} =2.6 E_{jet} T$ for the LBT model \cite{He:2015pra}. 

As follows from Fig. \ref{fig:dE_p}, the energy loss $dE/dx$ from the DQPM with $\alpha_{jet}=0.3$ and the DQPM in the pQCD limit show a similar behavior to the pQCD models BAMPS and LBT, while the results of the DQPM with both $\alpha(T)$ vertices show a faster growing with energy at low $T$ due to the rise of $\alpha_S$ near $T_C$. This behavior is similar to $\hat q/T^3$ in Fig. \ref{fig:qhat_p}.

The deviation at larger $T$ is related to different regularization schemes and Debye masses used in the calculations. Without showing explicitly, we mention that the drag coefficient $\mathcal{A}$ shows a very similar behavior - the difference between the drag coefficient $\mathcal{A}$ and the energy loss $dE/dx$ is almost negligible since for a very energetic light parton $\Delta E \approx |\Delta \vec{p}|$. We note that a similar behavior of the drag and diffusion coefficients, i.e. a small increase with $T$ and a strong rise with $p$, have been observed for charm quarks in Ref. \cite{Berrehrah:2014kba}.


\section{Summary}
\label{sec:summary}

We have studied the energy loss of fast jet partons by elastic scattering with off-shell quarks and gluons during the propagation through the strongly interacting quark-gluon plasma. The non-perturbative properties of the sQGP are described within the effective dynamical quasi-particle model (DQPM) which interprets the lQCD results on the QGP thermodynamics (energy density, pressure, susceptibility etc.) in terms of thermodynamics of off-shell quasiparticles with ($T,\mu_B$)-dependent masses and widths and broad spectral functions. In this respect, the DQPM differ from pQCD models dealing with on-shell massless pQCD partons and allows studying the non-perturbative features of QCD. The (quasi-)elastic interactions of off-shell partons in the DQPM are evaluated on the basis of the leading order Feynman diagrams with effective propagators and vertices.

Based on the DQPM, we estimated the jet transport coefficients: the transverse momentum transfer squared per unit length $\qhat$ as well as the energy loss per unit length $\Delta E =dE/dx$ in sQGP at finite $T$ and $\mu_B$. We stress that presently we discarded the radiative processes, i.e. gluon Bremsstrahlung.

We investigated the dependence of the transport coefficients on the temperature $T$, the baryon chemical potential $\mu_B$ as well as on the jet properties such as the momentum of jet parton, mass, flavor, and the choice of the strong coupling constant. We compared our results with the pQCD results obtained by the BAMPS model as well as with other theoretical approaches such as lattice QCD and the LBT model with the LO-HTL Debye masses and also with estimates of $\hat q$ by the JET and JETSCAPE Collaborations based on a comparison of hydrodynamical calculations with experimental heavy-ion data.
 
We summarize our findings as follows:
\begin{itemize}
\item The jet transport coefficients obtained in the pQCD limit of the DQPM (on-shell partons with zero masses and widths, constant $\alpha_S=0.3$) reproduce the results of the pQCD models such as BAMPS and the LBT model when choosing a similar Debye mass for the regularization of infrared divergences.

\item We obtained a strong increase of $\qhat/T^3$ when approaching $T_C$ which is attributed to the rise of the strong coupling $\alpha_S(T,\mu_B)$ at $T\to T_C$. When replacing $\alpha_S \to 0.3$ at the jet parton vertex the temperature dependence of $\qhat/T^3$ is getting weaker, which is even more visible in the pQCD limit. Moreover, $\qhat$ strongly increases with the momentum of jets.

\item The scaled energy loss per unit length $(1/T^2) dE/dx$ (as well as the drag $\mathcal{A}$ coefficient) show an increase with temperature $E/T$ (similar to $\qhat/T^3$) which is larger in the DQPM at $T \to T_C$ due to the rise of $\alpha_S(T,\mu_B)$. 

\item We have estimated the $\mu_B$ dependence of the transport coefficients and found the decrease of $\qhat$ with $\mu_B$ due to a reduction of the cross sections with $\mu_B$. The latter might be of interest for the interpretation of jet attenuation at SPS and low RHIC energies where the baryon chemical potential is non-zero.

\item We studied the influence of the off-shellness of the medium partons on $\qhat$ (i.e. computed within Eq. (\ref{eq:O_off}) by comparing with the on-shell evaluation of $\qhat$ within Eq. (\ref{eq:O_on}) assuming the pole mass of medium partons - cf. Fig. \ref{fig:qhat-T_OnOff}. We have obtained that the off-shellness reduces $\qhat$ especially at low $T$.

\item We indicate that the suppression of a gluon jet is stronger than that of quark jet and increases with $T$ mainly due to the color factor as well as a non-trivial $T$-dependence of quark and gluon cross sections. Also we have shown that the largest energy loss of quark and gluon jets stems from the interaction with medium gluons due to the larger $gg$ cross sections compared to $qq$ or $qg$ cross sections.

\item The comparison of our results with pQCD models - BAMPS and LBT - shows that the energy loss of a jet parton in the non-perturbative QCD medium (as characterized by the DQPM) occurs much more effectively than by scattering with massless pQCD partons. Furthermore, at large $T$ our results for $\qhat$ are in qualitative agreement with pQCD results, with lattice results (pure SU(3) and (2+1)-flavor) as well as with phenomenological estimates by the JET and JETSCAPE collaborations and Color String Percolation Model. However, we note that for a quantitative comparison with phenomenologically extracted $\qhat$ from a fit of jet observables measured in heavy-ion experiments, we need to account for the radiative energy loss, too, which is a subject of an upcoming study.
\end{itemize}

Thus, our study of jet transport coefficients shows a large sensitivity of the jet energy loss to the properties of the QCD medium: weakly interacting pQCD versus the strongly interacting non-perturbative QGP.


\section*{ACKNOWLEDGEMENTS}

The authors acknowledge inspiring discussions with J. Aichelin, W. Cassing, M. Djordjevic, P.-B. Gossiaux, H. van Hees, G. Moore and F. Senzel. We thank to A. N. Mishra, D. Sahu, and R. Sahoo for providing us the result from the CSPM in data format. Furthermore, we acknowledge support by the Deutsche Forschungsgemeinschaft (DFG, German Research Foundation) through the grant CRC-TR 211 "Strong-interaction matter under extreme conditions" - project number 315477589 - TRR 211. I.G. also acknowledges support from the "Helmholtz Graduate School for Heavy Ion research". The computational resources have been provided by the Goethe-HLR Center for Scientific Computing.


\bibliography{refs}

\end{document}